\documentclass{optica-article}
\hypersetup{colorlinks=true,linkcolor=blue,citecolor=blue}
\journal{opticajournal} 

\articletype{Research Article}
\usepackage{lineno}
\usepackage{subcaption}

\begin{document}

\title{\textbf{Single-photon generation at room temperature using molecular optomechanics in a hybrid photonic-plasmonic cavity}}

\author{Shabnam Abutalebi B.A.\authormark{1,*}, Seyed Mahmoud Ashrafi\authormark{2,3}, Hassan RanjbarAskari\authormark{1}, Alireza Bahrampour\authormark{2,3} }

\address{\authormark{1} Department of Physics, Faculty of Science, Vali-e-Asr University of Rafsanjan, Rafsanjan, Iran\\
\authormark{2} Department of Physics, Sharif University of Technology, Tehran, Iran\\
\authormark{3} Center for Quantum Engineering and Photonics Technology,
Sharif University of Technology, Tehran, Iran}

\email{\authormark{*}shabnam.abutalebi@stu.vru.ac.ir} 


\begin{abstract*} 
 We propose a novel integrated structure for single photon generation at room temperature based on a molecular optomechanics system in a hybrid photonic-plasmonic cavity.  The proposed structure comprises a single molecule within a plasmonic cavity, coupled to a 2D photonic crystal resonator. In this paper, we theoretically identify the ability of the scheme through calculation second order correlation function $g^2 (0)$ for four different coupling regimes. We demonstrate the quantum paths and the destructive interference mechanism through the selection of efficient and preferred basis. Furthermore, we find that the unconventional photon blockade effects can occurs in the weak molecular optomechanics coupling. This structure holds the potential to serve as an integrated single-photon source for quantum networks at room temperature.
\end{abstract*}

\section{Introduction}
In the field of quantum optics, where the manipulation of individual photons and their interactions holds the key to quantum information processing and communication, the concept of photon blockade (PB) emerges as a pivotal phenomenon \cite{luo2021study,ren2021antibunched,flamini2018photonic}. This effect pertains to the strong suppression of subsequent photon emissions following the initial emission of a single photon, fundamentally altering the probabilistic nature of photon emission processes \cite{lin2020kerr,ren2022antibunched}. Second-order correlation function ($g^{(2)}$) is key for single-photon source quality, indicating multi-photon emission probability and mean-variance relationship. If variance is less than mean ($g^{(2)} <1$), indicating sub-Poissonian behavior \cite{migdall2013single}.
Until now, three photon blockade with different physical mechanisems have been arisen. first, conventional photon blockade (CPB) arises from anharmonicity, where nonlinearity in the system prevents the absorption of a second photon with a specific frequency. This mechanism require strong coupling regime and explored in various systems, often involves Kerr-type interactions or atom-resonator couplings \cite{majumdar2013single, lin2020kerr}.
   Second, unconventional photon blockade (UCPB) arises due to quantum destructive interference and excels in situations characterized by extremely low mean photon numbers \cite{abo2022hybrid, xia2022improvement, snijders2018observation}. While UCPB may reduce the probability of generating a single photon, its ability to suppress multi-photon states and induce higher-order coherence presents  both a challenge and an advantage in the context of single-photon source applications.
   The last, non-Hermitian photon blockade (NHPB) mechanism occurs when there is a significant difference in the losses between the singly and doubly excited states of the system \cite{ben2023non}. This mechanism is not constrained by linewidths in contrast to conventional photon blockade.\\
Photon blockade has been predicted in various optomechanical systems, including Fabry-Perot cavities \cite{wang2020enhanced}, microtoroids \cite{wang2020photon} and photonic crystal cavities \cite{flayac2015all}. 
Recently, a new optomechanical approach has been introduced for surface-enhanced Raman scattering (SERS) in which there is an interesting optomechanical coupling between molecular vibrations and localized surface plasmon resonance (LSPR) mode similar to conventional cavity optomechanical systems \cite{roelli2016molecular,shlesinger2021integrated}. These new approaches could perfectly demonstrate some unknown observation in experiments like back-action amplification of Raman enhancement in the molecular samples \cite{esteban2022molecular}.
In molecular optomechanics, molecular vibrations serve as some of the smallest mechanical oscillators amenable to nano engineering. The energy and strength of molecular oscillations depend on specific functional groups and their chemical and physical environments \cite{dezfouli2019molecular}.
An important advantage of utilizing molecules as mechanical oscillators in optomechanical systems is their resistance to temperature effects, a factor that previously constrained the practical application of optomechanical systems. To enhance the interaction between molecule and photons, plasmonic cavity with small mode volume has been employed. However, this cavity often has a low Q-factor. Combining plasmonic cavity with high Q-factor optical cavity has enabled the compensation of this limitation, making them valuable for nanoscale optical physics and integrated photonics applications \cite{ ashrafi2020long,yang2011hybrid, santhosh2016vacuum, zhang2020hybrid, xiao2012strongly, barreda2021hybrid}.\\
In this study, we proposed a hybrid plasmonic- photonics resonator containing a molecular optomechanics system and a 2D photonic crystal. The 2D photonic crystal, designed with periodic dielectric structures exhibiting photonic bandgap properties, provide high-quality resonance. This hybrid configuration, which confines electromagnetic fields at subwavelength scales through LSPR cavity, enhances light-matter interaction.
On the other hand, plasmonic resonance supported by metallic nanocavity enable the boosting of Raman scattering from a single molecule. What sets our system apart is its operational viability at room temperature, eliminating the need for the extreme cooling methods often associated with traditional quantum systems. This characteristic not only simplifies experimental setups but also holds promise for practical, real-world applications.
The remaining part of the paper is organized as follows: In Sec. \ref{Proposed structure}, we introduce our proposed structure. The theoretical model of the proposed structure is presented in Sec. \ref{Theoretical model}. Numerical calculations and results can be found in Sec. \ref{Numerical method and results}. The paper concludes with a summary in Sec. \ref{Conclusion}.

\section{Proposed structure}\label{Proposed structure}

\begin{figure}[htbp]
	\centering
	\begin{subfigure}[b]{0.5\textwidth}
		\centering
		\includegraphics[width=1.2\textwidth]{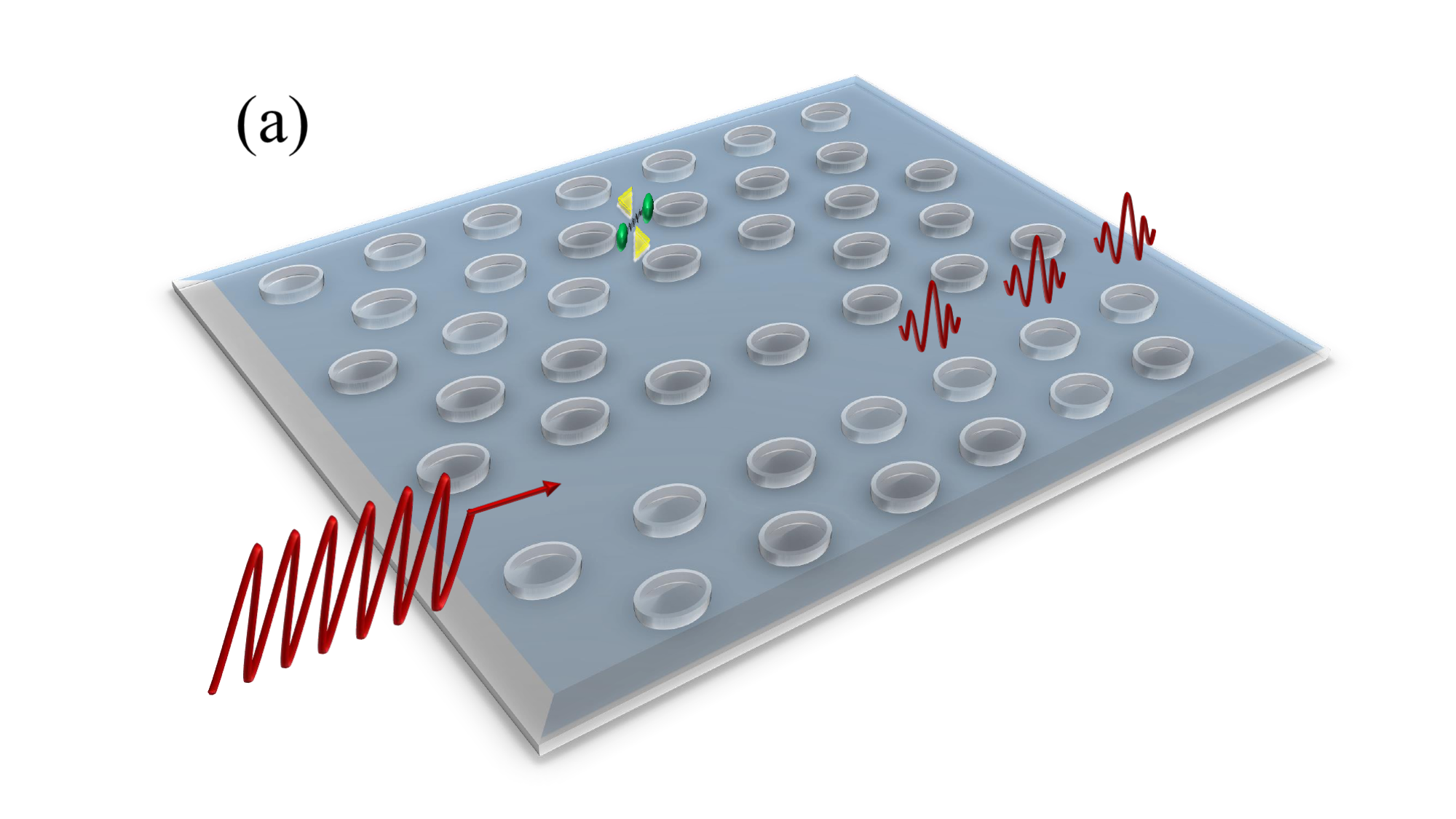}
	\end{subfigure}
	\hfil
	\begin{subfigure}[b]{0.3\textwidth}
		\centering
		\includegraphics[width=0.8\textwidth]{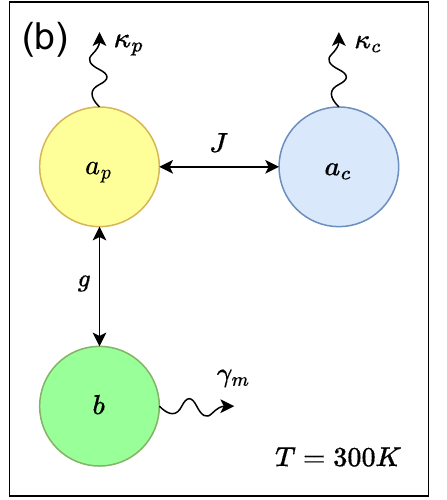}
	\end{subfigure}
	\caption{(a) Schematic diagram of the photonic crystal cavity coupled to a plasmonic nano-antenna cavity with a molecule positioned in the hot spot of the nano-antenna to form a molecular optomechanical system. (b) Interaction diagram of the cavities and molecule with damping rates. }
	\label{pic8}
\end{figure}

The primary objective of this study is to explore the capacity of a molecular optomechanical system to generate single photon at room temperature. We proposed a molecular optomechanical system consists of a Raman active molecule situated in the hot spot of the bowtie nano-antenna (BNA) coupled to photonic crystal (PC) resonator.  BNA is highly advantageous for confining light within ultra-small volume, surpassing the limits of diffraction but it has low quality factor.  For compensation of this limitation, the hybrid PC-BNA nano-resonator employed.  The PC has a high quality factor and is pumped by a laser with frequency $\omega_l$ and amplitude $\Omega$. Additionally, there is a waveguide for driving the cavity and the path of a single photon that is generated. This system shown in Fig. \ref{pic8}(a), and the interaction diagram of this system is presented in Fig. \ref{pic8}(b).
\\
\section{Theoretical model}\label{Theoretical model}
The Hamiltonian for the PC and BNA modes with central frequencies $\omega_c$ and $\omega_p$ is expressed as $H_0 = \sum_{i=c,p} \omega_i a_i^\dagger a_i$, where $a_c^\dagger (a_c)$ and $a_p^\dagger (a_p)$ represent the bosonic creation (annihilation) operators for the optical and LSPR modes, respectively. The system of units is chosen such that $\hbar=1$.
The molecule can also be treated as quantum harmonic oscillator with the frequency $\omega_m$. Consequently, its corresponding Hamiltonian is $H_m = \omega_m b^\dagger b$, where $b^\dagger (b)$ is the bosonic creation (annihilation) operator for the vibrational mode. It is worth to mention that the primary assumption for coherent coupling between mechanical oscillator and LSPR mode is $(\omega_p \gg \omega_m)$ \cite{roelli2016molecular}. The corresponding molecular optomechanics interaction can be written as $H_{\text{BNA-m}}=g a_p^\dagger a_p (b^\dagger + b)$, in which the optomechanical coupling factor is $g = \frac{\omega_{p}}{\epsilon_0 V_p} \frac{1}{\sqrt{ R_m /2\omega_{m}}}$, with $R_m$ representing the Raman activity associated with the vibration under study and $V_p$ being the effective mode volume of the plasmonic cavity mode \cite{dezfouli2019molecular}. Due to the large mode volume of the PC cavity, we can neglect the optomechanical interaction between the PC mode and the molecule.
Meanwhile, we consider an interaction between the non-local optical mode in the PC and LSPR modes by the Hamiltonian $H_{\text{PC-BNA}}=J(a_c^\dagger a_p + a_p^\dagger a_c)$. Here, $J$ is the coupling strength between the PC and LSPR modes, strongly dependent on the distance between the nano-antenna and the PC cavity \cite{dezfouli2017modal}.
By performing a rotating transformation defined as $
U = \exp\left[-i(\omega_c a^\dagger_ca_c +\omega_p a^\dagger_pa_p)t\right],
$
the system Hamiltonian is transformed as:
\begin{align}
	H = \Delta_c a_c^\dagger a_c + \Delta_p a_p^\dagger a_p + \omega_m b^\dagger b+J(a_c^\dagger a_p + a_p^\dagger a_c)- g a_p^\dagger a_p (b^\dagger + b)+ \Omega(a_c^\dagger + a_c).\label{eq1}
\end{align}
Here, $\Delta_{c}=\omega_{c}-\omega_l$ and $\Delta_{p}=\omega_{p}-\omega_l$ represent the detuning of the photonic and plasmonic cavities from the pump frequency, respectively.
The first three terms correspond to the optical cavity, plasmon cavity, and phonons of molecular vibration. The eigenvalues of these terms are harmonic, and their eigenvectors are denoted as $|{nmq} \rangle $ (bare basis), where n, m, and q represent the photon number in the optical cavity, the plasmon number, and the phonon number of molecular vibration, respectively.
The fourth term accounts for the photon-plasmon cavity coupling. Since this term is bilinear with respect to annihilation and creation operators, its eigenvalues are also harmonic and do not induce the CPB effect. The corresponding eigenvectors are denoted as $|n_{+}n_{-} q\rangle$, where $n_{+}$ and $n_{-}$ represent the number of particles associated with resonance frequencies $\omega_{+}$ and $\omega_{-}$, respectively. The $\omega_{+}$ and $\omega_{-}$ are introduced in Sec. \ref{The nonHermitian Schrödinger equation}.
The fifth term is non-harmonic and represents the interaction between plasmons and phonons. This non-harmonic term results in non-harmonic eigenvalues for the first five terms of the Hamiltonian. The eigenvectors for the first five terms of the Hamiltonian are denoted as $|E_{nmq}\rangle$.
This term introduces nonlinear contributions in the Heisenberg-Langevin equation and leads to the generation of new frequencies, known as the Stokes and anti-Stokes frequencies of different orders. These frequencies can be used for cooling and warming of cavity modes.
The last term represents the laser input, and its effect is to displace the eigenvectors and eigenvalues.\\
In this work, we analyze the statistical properties of photons using the second-order correlation function at zero delay time.
\begin{align}
	g^{(2)}(0) = \frac{{\langle  a^\dagger a^\dagger a a\rangle}}{{\langle  a^\dagger a\rangle^2}} \label{eq2}
\end{align}
$a$ and $a^\dagger$ are the annihilation and creation operators of the mode under investigation. The value of $g^{(2)}(0)<1$ corresponds to the single photon statistics. We utilize both the Lindblad equation and the Schrödinger equation with non-Hermitian Hamiltonian methods. To ensure the validity of our findings, we compare the results obtained from these two approaches.
To comprehensively characterize the dynamical behavior of the system, it is crucial to consider the interactions between phonons, plasmons, photons modes, and their respective environments. 
\subsection{The Lindblad equation}
The decoherence of the system can be described by the evolution of the density matrix using the Lindblad master equation \cite{manzano2020short}:
\begin{align}
	\nonumber&\frac{{d \rho}}{{dt}} = -i [H, \rho] + \mathcal{L}_{a_c} + \mathcal{L}_{a_p} + \mathcal{L}_{b}\\
	&\mathcal{L}_{a_i} = \frac{\kappa_i}{2} \mathcal{D}_{a_i}[\rho], \quad i = c,p\\
	\nonumber&\mathcal{L}_{b} = \frac{\gamma_m}{2} ( n_{th}+1) \mathcal{D}_{b}[\rho] + \frac{\gamma_m n_{th}}{2} \mathcal{D}_{b^\dagger}[\rho],\label{eq8}
\end{align}
where $\rho$ is the dynamical density matrix of the system, which can reach its own steady state $\rho_s$ after a long evolution time. $n_{th}$, $\kappa_i$, and $\gamma_m$ are the mean phonon number of the molecule (which depends on the temperature of its heat bath), the optical cavity field decay (PC and Plasmon cavity), and the molecular damping rate, respectively.
The term $\mathcal{D}_{C}[\rho]$ is referred to as a Lindblad superoperator, and it can be written as follows:
\begin{align}
	\mathcal{D}_{C}[\rho] = 2C \rho C^\dagger - C^\dagger C\rho - \rho C^\dagger C.
\end{align}
Eq. (\ref{eq2}) is employed to analyze the photon blockade mechanism.
\subsection{The non-Hermitian Schrödinger equation}\label{The nonHermitian Schrödinger equation}
To observe the dynamical behavior in our system, the non-Hermitian Schrödinger equation can be employed. The non-Hermitian Hamiltonian is constructed by introducing a phenomenological imaginary dissipative term into the effective Hamiltonian:
\begin{align}
	\hat{H'} = \hat{H}- \frac{i}{2}(\kappa_c \hat{a}_c^\dagger \hat{a}_c + \kappa_p \hat{a}_p^\dagger \hat{a}_p + \gamma_m \hat{b}^\dagger \hat{b}).\label{eq3}
\end{align}
Where, $\kappa_{c}$, $\kappa_{p}$ are decay rate of photonic cavity, plasmonic cavity and $\gamma_m$ is the molecular damping rate. The term $g a_p^\dagger a_p (b^\dagger + b)$ in the Hermitian part of the effective Hamiltonian introduces anharmonicity in the energy spectrum of the system. The degree of anharmonicity depends on the coupling coefficient $g$, with higher values of $g$ leading to greater anharmonicity and a more pronounced CPB effect. In the absence of the plasmon-phonon coupling term or when $g$ has a small value, the CPB effect is not observed.
To investigate the UCPB effect, it's crucial to consider the quantum interference paths. For this purpose, we obtain the solution of the non-Hermitian Schrödinger equation. The Schrödinger equation can be expressed in an arbitrary basis, especially when the Hilbert space has infinite dimensions. This transformation results in a system of infinite ordinary differential equations with an infinite number of unknowns. To solve this system of infinite differential equations, we can employ a truncation method. The convergence speed of this method is influenced by the choice of basis. A basis that yields fast convergence is often referred to as an efficient basis which is similar to optimal wavelets in signal analysis \cite{GUPTA2005147}. Depending on the coupling coefficients $g$ and the strength of $J$, different preferred bases can be employed, so that four different regime are arrived as follow:\\
i) Both $J$ and $g$ are relatively small compared to the cavity decay rate. The preferred basis consists of the eigenvectors of the non-interacting part of the Hermitian Hamiltonian, denoted as $\{|nmq \rangle \}$. In this case, the eigenvalues are represented as $E_{nmq} = n\Delta_{c} + m\Delta_{p} + q\omega_{m}$.
The quantum state $\left|\psi\right\rangle = \sum C_{nmq}|nmq\rangle $ is expressed in this basis, and the Schrödinger equation takes the form:
\begin{align}
	\begin{split}
		i\dot{C}_{jkl} & = E^\prime_{jkl} C_{jkl} + J \sum_{n,m,q} \left[ \sqrt{n(m+1)} C_{n-1,m+1,q} + \sqrt{m(n+1)} C_{n+1,m-1,q} \right] \\
		& - g \sum_{n,m,q} \left[ m\sqrt{q+1} C_{n,m,q+1} + m\sqrt{q} C_{n,m,q-1} \right] \\
		& + \Omega \sum_{n,m,q} \left[ \sqrt{n} C_{n-1,m,q} + \sqrt{n+1} C_{n+1,m,q} \right] ;\qquad j,k,l=0,1,2,...
	\end{split}\label{eq4}
\end{align}
Where
$E^\prime_{jkl} = E_{jkl} - \frac{i}{2} \left( j\kappa_c + k \kappa_p + l \gamma_m \right)$
are the effective Hamiltonian eigenvalues in the absence of couplings. The eigenvalues $E_{jkl}=j \Delta_{c}+k\Delta_{p}+l\omega_{m}$ are obtained from the Hamiltonian in the absence of loss terms. The system in Eq. (\ref{eq4}) represents a system of infinite ordinary differential equations with an infinite number of unknowns.
In the absence of laser input ($\Omega$=0) and coupling terms, the sum of the photon and plasmon number operators commutes with the effective Hamiltonian, and $N=N_c+N_p$ is a constant of motion. The anharmonicity term facilitates energy transfer between phonons and plasmons through different orders of Stokes and anti-Stokes effects, leading to changes in the number of plasmons during the system's evolution. In the weak coupling regime, these effects are negligible. Furthermore, it is assumed that the bandwidth of the cavities is so small that neither the Stokes nor the anti-Stokes frequencies can excite the cavity modes.
The photon-plasmon coupling does not alter the harmonic behavior of the system but transforms the phonon-plasmon dynamics into positive and negative polariton time evolution. For small values of $J$, it does not have a significant effect on the conservative observables. However, the laser input intensity changes the photon and plasmon numbers of the system, making the conservation of $N$ ineffective for reducing the dimension of the Hilbert space. For low-intensity laser input ($\Omega\ll\kappa_c$), the total number of photons and plasmons is limited from above, allowing for the truncation of the system of governing equations.\\
 For weak laser input and a small coupling coefficient $g$, the total number of high-energy particles (photons and plasmons) remains constant, and the basis $\{ |p,q\rangle; p=n+m \}$ is efficient for our analysis. The maximum number of high-energy particles is denoted by p, and the vector space spanned by the eigenvectors of the Hamiltonian in the absence of input and optomechanical coupling coefficient, with $p_i$ high-energy and q low-energy particles, is denoted as $V_{p_i}$. Therefore, $H = \bigoplus_{p_i \le p} V_{p_i}$ is an efficient Hilbert space for the UCPB effect. $\bigoplus_{p_i \le p}V_{p_i}$ represents the direct sum of vector spaces $V_{p_i}$.\\
 ii) When the coupling coefficient between the optical and plasmonic cavities $J$ is negligible, but the phonon-plasmon coupling coefficient $g$ is not negligible, the effect of $g$ on anharmonicity is significant. However, its effect on the photon and plasmon numbers is negligible, making the basis $\{ |nmq\rangle \}$ an efficient choice for truncating the governing equations.\\
iii) When the coupling coefficient $J$ is not negligible, but $g$ is small. The transformation $a_{+}=a_c \cos \theta +a_p \sin \theta $ and $a_{-} =- a_c \sin \theta +a_p \cos \theta $ ($\tan{2\theta}=\frac{2J}{\Delta_{c}-\Delta_{p}}$) can be used to diagonalize the cavities part of the Hamiltonian as follows:
\begin{align}
	\begin{split}
H^{\prime\prime} &=\Delta_{+}a^{\dagger}_{+}a_{+}+\Delta_{-}a^{\dagger}_{-}a_{-}+\omega_m b^{\dagger}b-g_{+}a^{\dagger}_{+}a_{+}(b^{\dagger}+b)+-g_{-}a^{\dagger}_{-}a_{-}(b^{\dagger}+b)\\&-g_{\pm}(a^{\dagger}_{+}a_{-}+a^{\dagger}_{-}a_{+})(b^{\dagger}+b)+i \Omega _{+}(a^{\dagger}_{+}+a_{+})+i \Omega_{-}(a^{\dagger}_{-}+a_{-}).\label{eq5}
	\end{split}
\end{align}
In this context, the coupling strength is defined as follows: $g_{+} = g\cos^2\theta$, $g_{-} = g\sin^2\theta$, and $g_{\pm} = g\sin\theta\cos\theta$. Additionally, the drive of the system is given by $\Omega_{+} = -\Omega\sin\theta$ and $\Omega_{-} = \Omega\cos\theta$. Here, the preferred basis consists of the eigenvectors of the Hamiltonian in the absence of anharmonicity and laser input, denoted as $\{ |n_{+}n_{-}q\rangle $: $n_{+}$ and $n_{-}$ represent the number of particles with $\omega_{+}$ and $\omega_{-}$ frequencies, respectively$\}$. The detuning frequencies $\Delta_{\pm}=\omega_{\pm}-\omega_l$ can be determined using the following relation:
\begin{align}
\Delta_{\pm}=\frac{1}{2}(\Delta_{c}+\Delta_{p})\pm \frac{1}{2}\sqrt{(\Delta_{c}-\Delta_{p})^2+4J^2}.\label{eq6}
\end{align}
When both $\Delta_{c}$ and $\Delta_{p}$ are complex, $\Delta_{+}$ and $\Delta_{-}$ are also complex and are denoted by $\Delta_{+}'$ and $\Delta_{-}'$, by replacing $\Delta_{c}$ and $\Delta_{p}$ with their complex values $\Delta_{c}' = \Delta_{c} - i\kappa_{c}$ and $\Delta_{p}' = \Delta_{p} - i\kappa_{p}$ in Eq. (\ref{eq6}). The complex part of $\Delta_{+}'$ is near $\kappa_{p}$, while the complex part of $\Delta_{-}'$ is close to $\kappa_{c}$, hence the particles with $\omega_{+}$ and $\omega_{-}$ frequencies are called plasmon-like and photon-like particles. The state vector $\left|\psi\right\rangle =\sum U_{n_{+}n_{-}q}|n_{+}n_{-}q\rangle $ is written in the new basis and is substituted to the non-Hermitian Schrödinger equation, the following equations are obtained:
\begin{align}
	\begin{split}
		i\dot{U}_{jkl} & = F^\prime_{jkl} U_{jkl} - g_{+} \sum_{n_{+},n_{+},q} \left[ n_{+}\sqrt{q+1} U_{n_{+},n_{-},q+1} + n_{+}\sqrt{q} U_{n_{+},n_{-},q-1} \right]\\&- g_{-} \sum_{n_{+},n_{-},q} \left[ n_{-}\sqrt{q+1} U_{n_{+},n_{-},q+1} + n_{-}\sqrt{q} U_{n_{+},n_{-},q-1} \right] \\
		&-g_{\pm} \sum_{n_{+},n_{+},q} [ \sqrt{(n_{+}+1) n_{-}}\sqrt{q+1} U_{n_{+}+1,n_{-}-1,q+1} +  \sqrt{(n_{+}+1) n_{-}}\sqrt{q} U_{n_{+}+1,n_{-}-1,q-1} \\
		& + \sqrt{n_{+} (n_{-}+1)}\sqrt{q+1} U_{n_{+}-1,n_{-}+1,q+1} +  \sqrt{n_{+} (n_{-}+1)}\sqrt{q} U_{n_{+}-1,n_{-}+1,q-1} ]\\&+ \Omega_{+} \sum_{n_{+},n_{-},q} \left[ \sqrt{n_{+}+1} U_{n_{+}+1,n_{-},q} + \sqrt{n_{+}} U_{n_{+}-1,n_{-},q}\right]\\&+ \Omega_{-} \sum_{n_{+},n_{-},q} \left[ \sqrt{n_{-}+1} U_{n_{+},n_{-}+1,q} + \sqrt{n_{-}} U_{n_{+},n_{-}-1,q} \right]; \qquad j,k,l=0,1,2,....
	\end{split}\label{eq7}
\end{align}
Where $F^\prime_{jkl} = F_{jkl} - \frac{i}{2} \left( j\kappa_{+} + k \kappa_{-} + l \gamma_m \right)$ are the effective Hamiltonian complex eigenvalues in the absence of couplings. Eigenvalues $F_{jkl}=j \Delta_{+}+k\Delta_{-}+l\omega_{m}$ are obtained from the Hamiltonian in the absence of loss terms. Eq. (\ref{eq7}) also represents a system of an infinite number of linear ordinary differential equations with an infinite number of unknowns. This system of linear ordinary differential equations can be solved using truncation methods. In the absence of excitation, the total number of photon-like and plasmon-like particles, denoted as $N_p= a^{\dagger}_{+}a_{+}+ a^{\dagger}_{-}a_{-}$, remains a constant of motion. In the presence of low input intensity ($\Omega_{\text{photon-like}}\ll \kappa_{\text{photon-like}}$), the number of particles is finite and can be estimated by the number of photons injected into the optical cavity.
With this initial estimation, we can employ the iteration method to determine a suitable truncation of the system of differential equations.\\
iv) When the effects of both coupling coefficients $J$ and $g$ are not negligible, the photon-plasmon coupling coefficient changes the behavior to that of photon-like and plasmonic-like particles. While $g$ affects harmonicity, it has a negligible effect on the total number of polaritons. In this case, the basis $\{ |n_{+}n_{-}q\rangle \}$ is efficient for numerical calculations.
This effective basis is independent of $g$ in high-quality cavities and can be chosen based on the coupling coefficient $J$ between the photon and plasmon cavities.

\section{Numerical method and results }\label{Numerical method and results}
\begin{figure}[h]
	\centering
	\begin{subfigure}[b]{0.5\textwidth}
		\centering
		\includegraphics[width=0.9\textwidth]{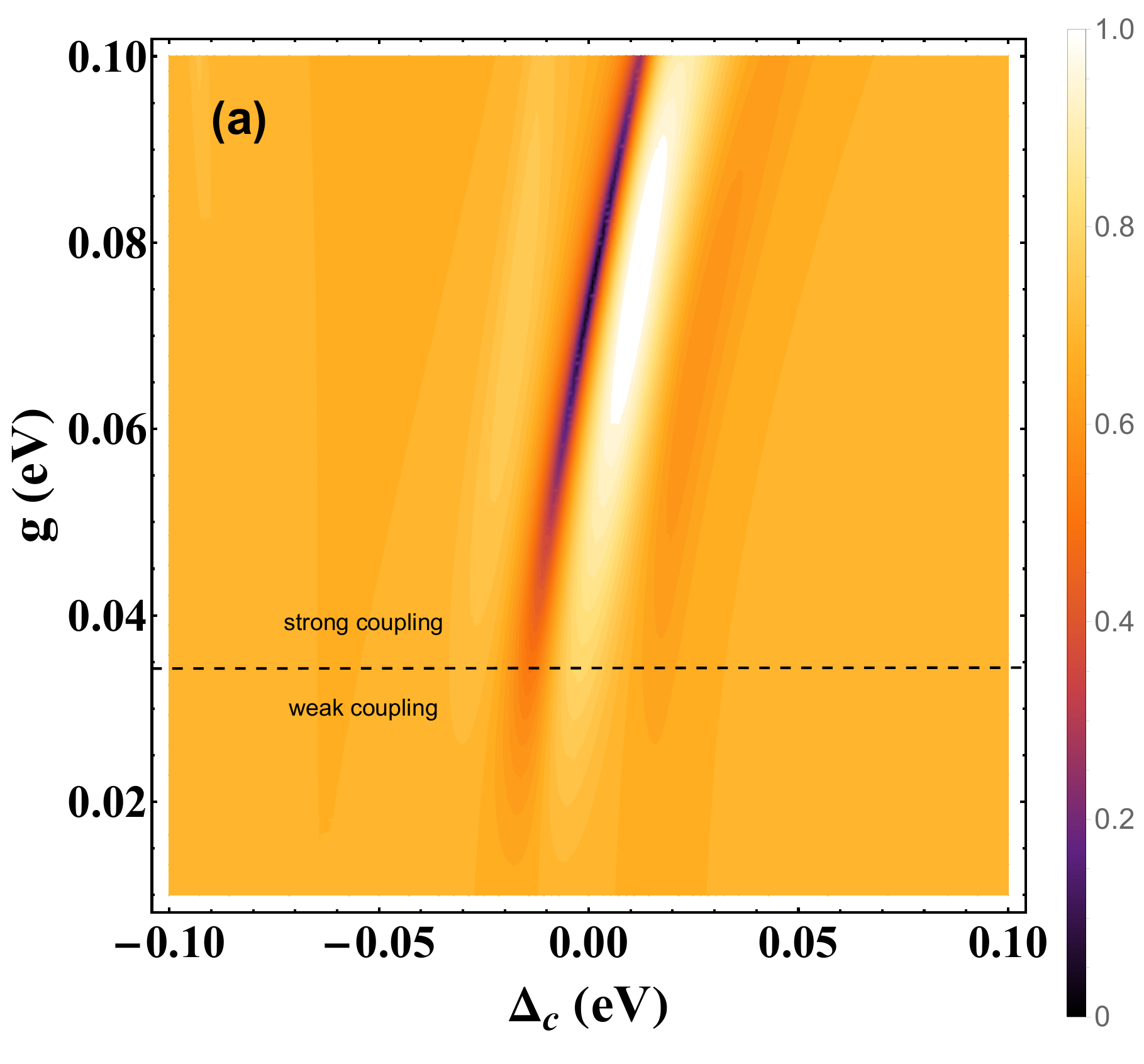}
		\caption{}
	\end{subfigure}
	\begin{subfigure}[b]{0.45\textwidth}
		\centering
		\includegraphics[width=\textwidth]{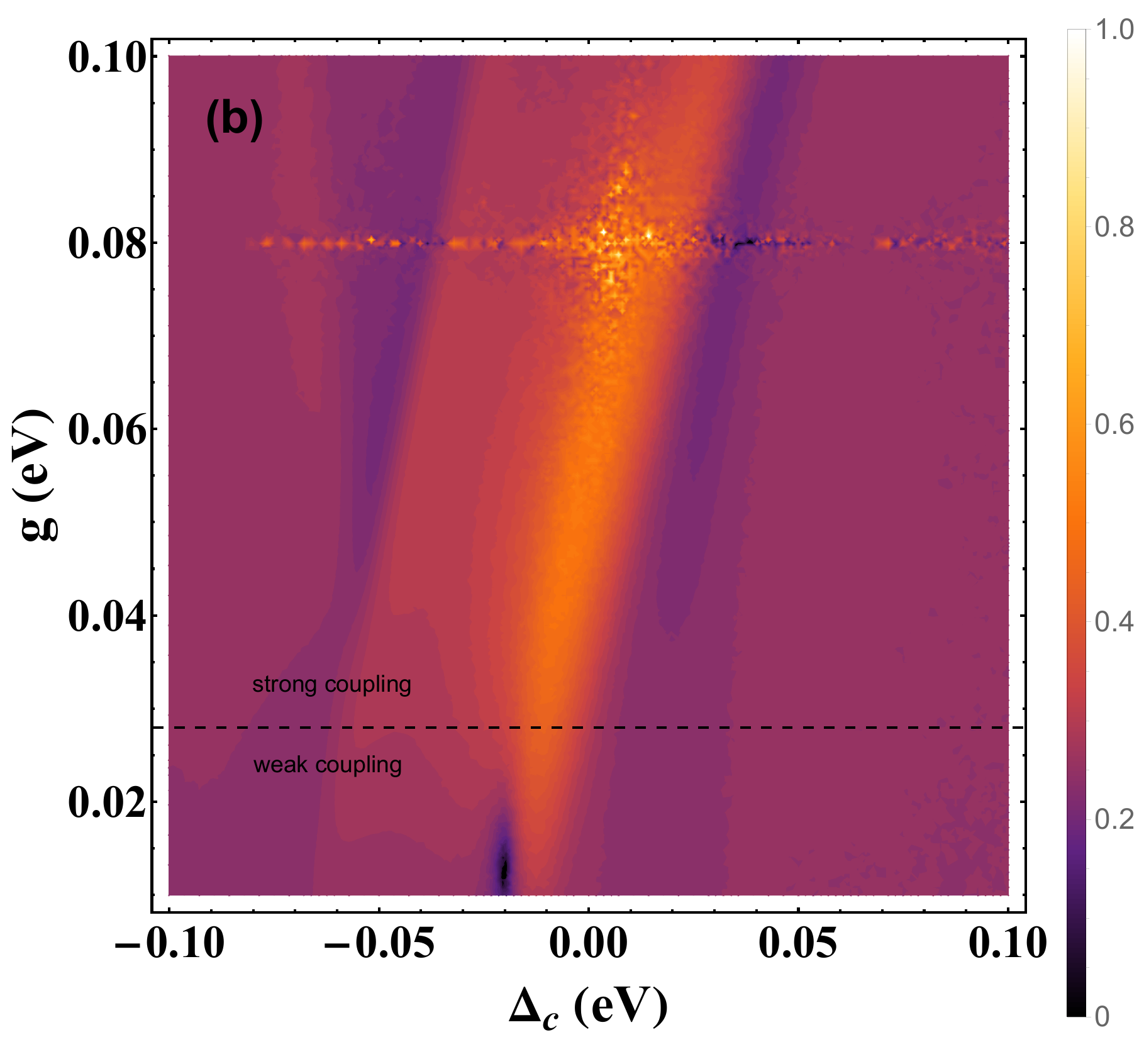}
		\caption{}
	\end{subfigure}
	\caption{The corerrelation function $g^{(2)}(0)$ versus BNA-molecule coupling coefficient $g$ and $\Delta_{c}$ with other parameters set to (a) $J=25 $ meV, $\kappa_c=1.342\times10^{-3}, \kappa_p=34$ meV,$\omega_{c}=1.342$ eV, $\omega_{p}=1.36$ eV, $\omega_{m}=200 $ meV and $ T=300$ K. (b) $J=40 $ eV, $\kappa_c=1.342\times10^{-3}, \kappa_p=27$ meV,$\omega_{c}=1.32 $ eV, $\omega_{p}=1.36$ eV, $\omega_{m}=150$ meV and $ T=300$ K.}\label{pic2}
\end{figure}

The system of governing Eqs. (\ref{eq4}) or (\ref{eq7}) in any basis is reduced to the following system of differential equations:
\begin{align}
	\dot{X}= AX, \label{eq10}
\end{align}
where $X$ is the vector of expansion coefficients of the state vector $\left|\psi\right\rangle$ and $A$ is the matrix of coefficients on the right hand side of governing equations. Dimensions of vector $X$ and  matrix $A$ is determined by the system parameters and chosen basis. The analytical solution of Eq. (\ref{eq10}) is given :
\begin{align}
	X(t)= \exp(At)X(0).\label{eq11}
\end{align}
To determine the dimension of matrix $A$ and obtain numerical results, the following algorithm is employed:\\
a) Based on the values of $J$, $\kappa_c$, and $\kappa_{p}$, select an efficient basis.\\
b) Estimate the number of high-energy particles (photons and plasmons) according to the input laser intensity.\\
c) Use room temperature to estimate the initial value for the phonon number ($q=q_0$).\\
d) Estimate the dimension of the truncated vector space.\\
e) Arrange the state vector in dictionary order.\\
f) Determine the elements of matrix $A$.\\
g) Solve the dynamic equation $\dot{X}=AX$ using analytical or numerical methods.\\
h) If $q=q_0$, store $X$ in register $X_0$ and increment $q$ by 1, then return to step $d$.\\
i) Compare $X$ and $X_0$ in the sense of the $\mathbb{C}^N$ metric. If the distance between $X$ and $X_0$, $d(X, X_0)$, is greater than a given $\epsilon$ (epsilon), replace $X_0$ with $X$, increment $q$ by 1, and return to step $d$.\\
j) If $N=N_0$, increment $N$ by 1 and return to step $d$.\\
k) If $d(X, X_0)$ is greater than $\epsilon$, replace $X_0$ with $X$, increment $N$ by 1, and return to step $d$.\\
l) Calculate $g^{(2)}$ and provide the values of $X$, $N$, and $q$.\\
 Our algorithm is employed for various values of the coupling coefficient $J$ and input intensity $\Omega$.
Our calculations show that for low input intensity ($\Omega \ll \kappa_c$), the speed of convergence is independent of the chosen basis. However, for high input intensity, the preferred basis depends on the value of $J$. For large values of $J$ ($J > \kappa_p$), the preferred basis is $\{ |n_{+}n_{-}q\rangle \}$, and for small values, no preferred basis exists.\\
For weak PC-BNA coupling strength (i and ii regimes), we consider a system consists of a PC cavity with a central frequency $\omega_{c}=1.342$ eV and a high quality factor, Q = $10^6$ (corresponding to $\kappa_c = 1.342\times10^{-3}$ meV). This PC cavity is coupled to a plasmonic nano-antenna with a central frequency, $\omega_p$, of 1.36 eV and a relatively low quality factor, $Q_p = 40$ (implying $\kappa_p = 34$ meV). A Raman-active molecule with a molecular vibrational energy of $\omega_m = 200$ meV and a damping rate of $\gamma_m = 0.2$ meV is placed in the hot spot of the nano-antenna to constitute a molecular optomechanics cavity setup. We also considered an external heat bath with a temperature of $T = 300$ K for the molecular system. Laser field coherently pumps the PC cavity, where the strength of the laser is taken to be weak ($\Omega$ = 0.01$\kappa_c$). We calculated $g^{(2)}(0)$ as a function of detuning $\Delta_{c}$ and optomechanical coupling $g$ for PC-BNA coupling strength $J=25$  meV ($J < \kappa_p$). To assess the influence of the basis on convergence speed, these calculations were performed in the bases: $\{ |nmq\rangle \}$ with the constraint $n+m=2$ (see \Ref {appendA}). As it shown in Fig. \ref{pic2}(a), the sub-Poissonian distribution
is achieved in the blue detuning and the specific domain of optomechanical coupling which divided in the weak and strong coupling regime.\\
For strong PC-BNA coupling strength (iii and iv regimes), we consider almost the same system with small different in plasmonic cavity ($\omega_p=1.32$ eV, $Q_p = 50$) and molecular frequency ($\omega_m = 150$ meV,$\gamma_m = 0.15$ meV). We calculated $g^{(2)}(0)$ as a function of detuning $\Delta_{c}$ and optomechanical coupling $g$ for PC-BNA coupling strength $J=40$  meV ($J > \kappa_p$). It is worth to mention that, these calculations were performed in the bases $\{ |n_{+}n_{-}q\rangle \}$ with the constraint $n_{+}+n_{-}=2$. As it shown in Fig. \ref{pic2}(b), there is a good region for single photon generation in weak coupling regime, because of a small $g^{(2)}(0) \approx 0.02$.  We analyze these results with their underlying physical mechanisms in the next two subsections. \\
\subsection{Weak coupling regime}
\begin{figure}[htbp]
	\centering
	\includegraphics[width=0.4\textwidth]{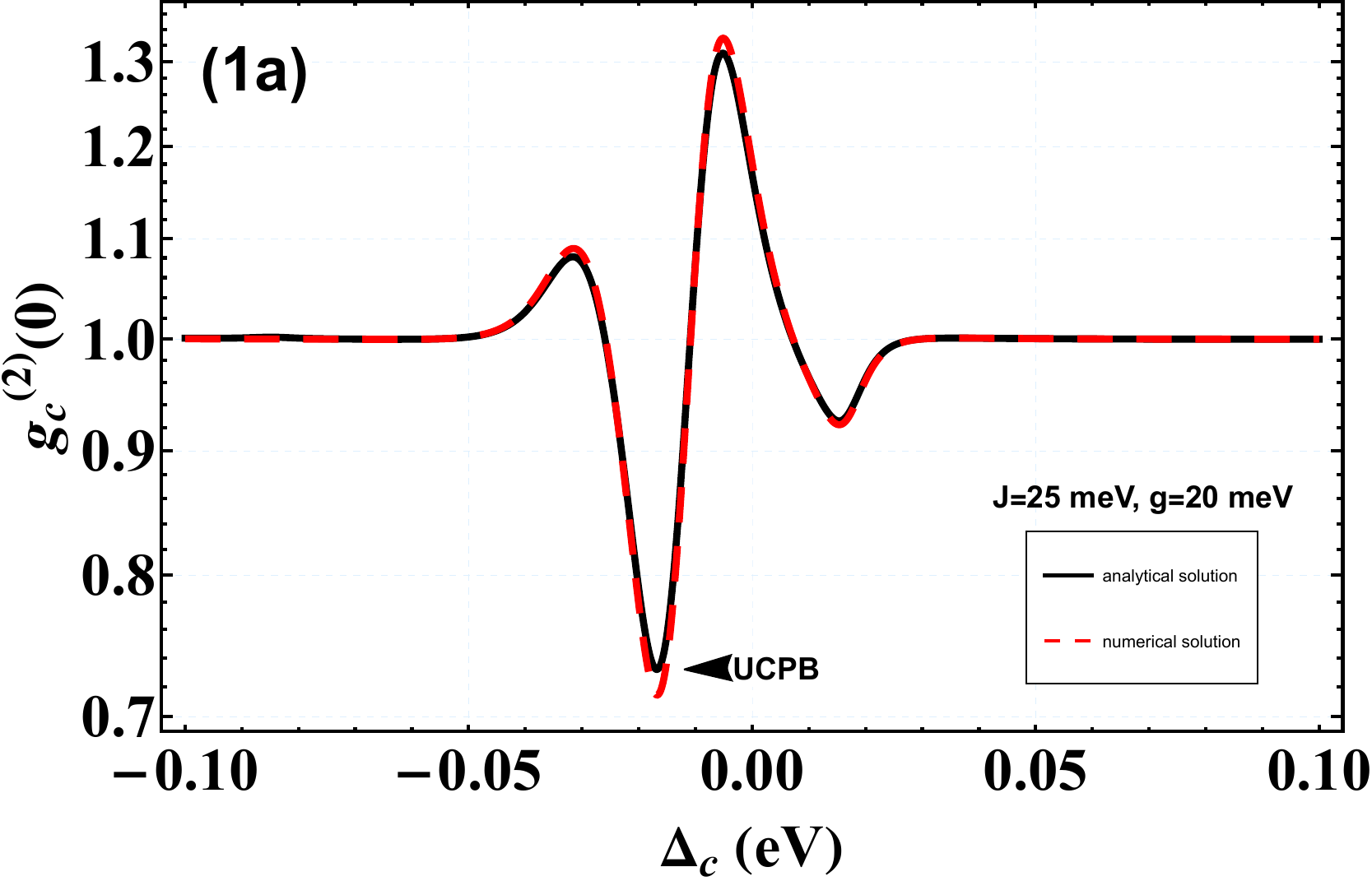}
	\hspace{1cm}
	\includegraphics[width=0.4\textwidth]{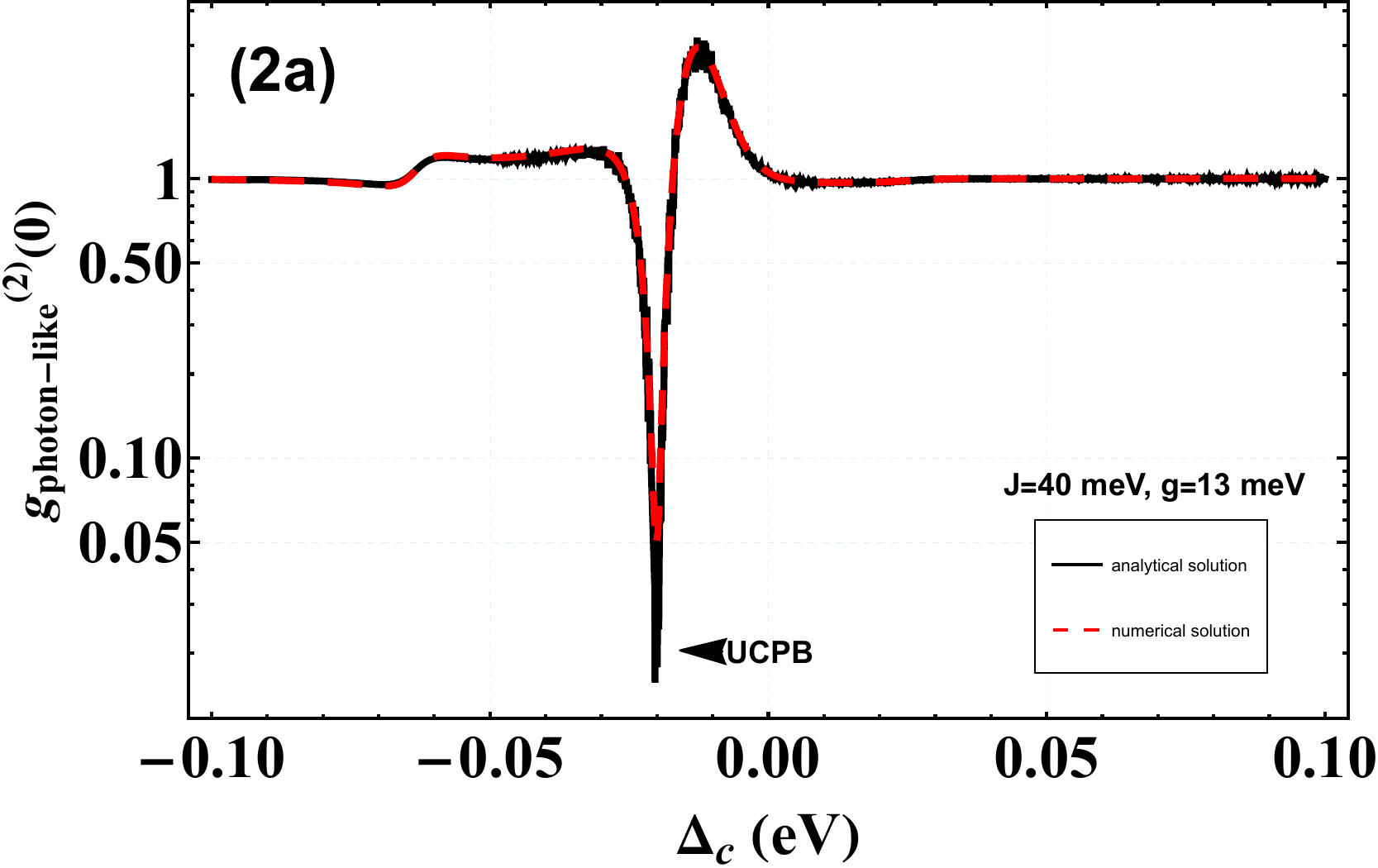}
	\hspace{1cm}
	\includegraphics[width=0.41\textwidth]{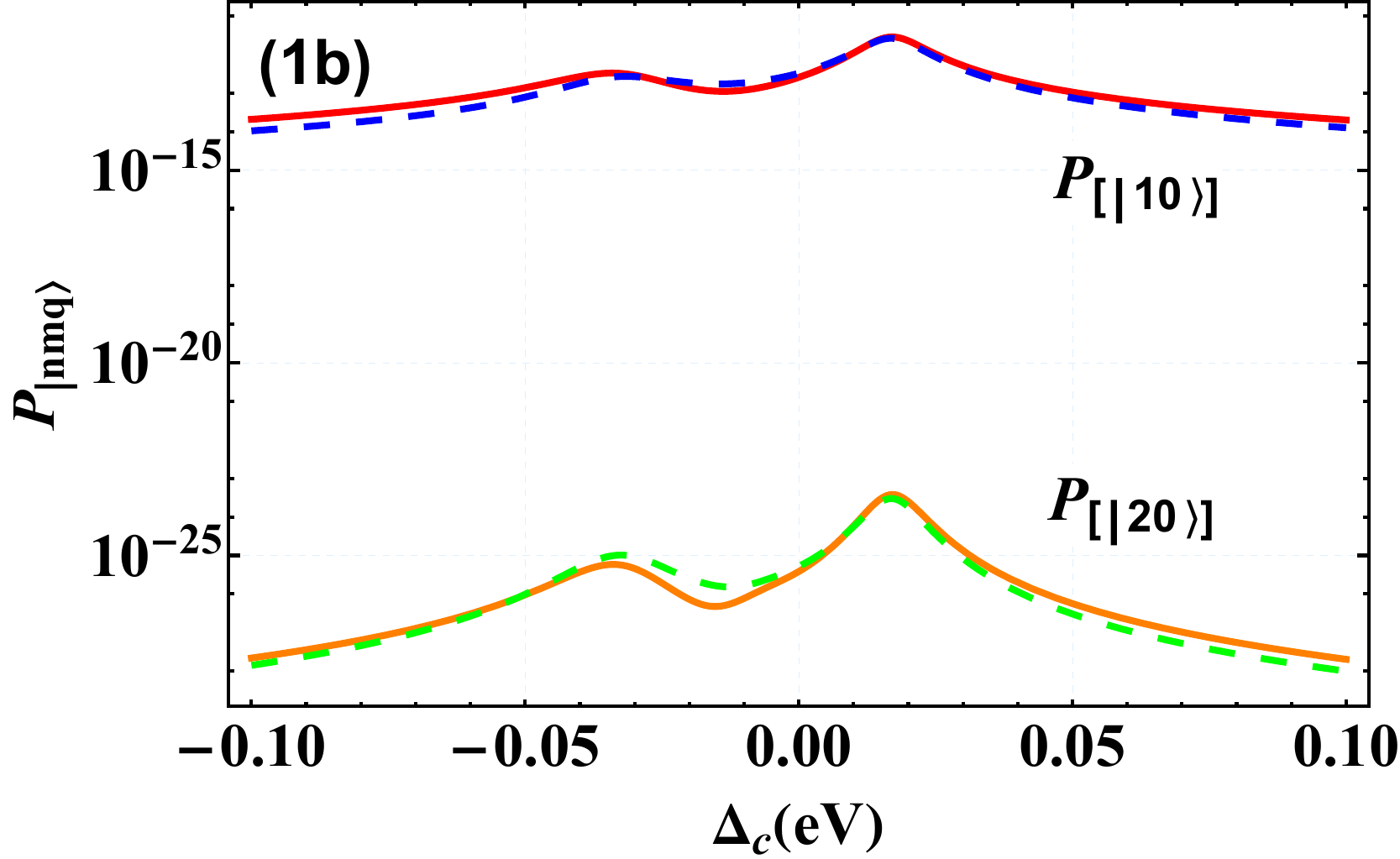}
	\hspace{0.95cm}
	\includegraphics[width=0.41\textwidth]{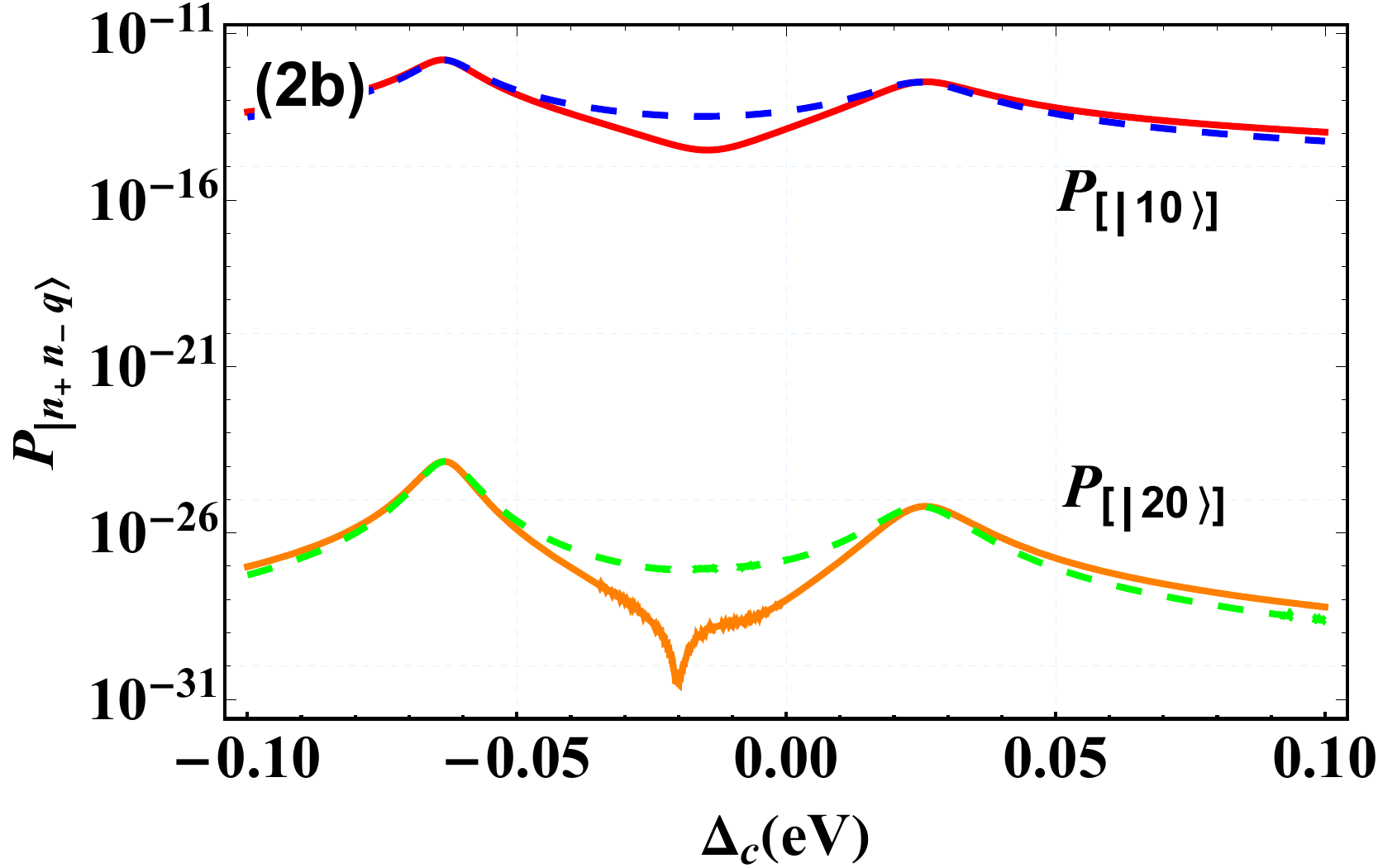}
	\caption{(1a) The correlation function $g^{(2)}(0)$ (1b) The state occupations $P_{[|10\rangle]}$ and $P_{[|20\rangle]}$ for the bare states versus the detuning $\Delta_{c}$ for $J=25$ meV and $g=20$ meV with other parameters set to $\kappa_c=1.342 \times 10^{-3}$ meV, $\kappa_p=34$ meV, $\gamma_m=200$ meV, $\omega_{c}=1.342$ eV, $\omega_{p}=1.36$ eV (3a) The correlation function $g^{(2)}(0)$. (2a) The correlation function $g^{(2)}(0)$ (2b) The state occupation probabilities $P_{[|10\rangle]}$ and $P_{[|20\rangle]}$ for the bare states versus the detuning $\Delta_{c}$ for $J=40$ meV and $g=13$ meV with other parameters set to $\kappa_c=1.32 \times 10^{-3}$ meV, $\kappa_p=27$ meV, $\gamma_m=150$ meV, $\omega_{c}=1.32$ eV, $\omega_{p}=1.36$ eV.}
	\label{pic12}
\end{figure}

\begin{figure}[htbp]
	\centering
		\includegraphics[width=0.6\textwidth]{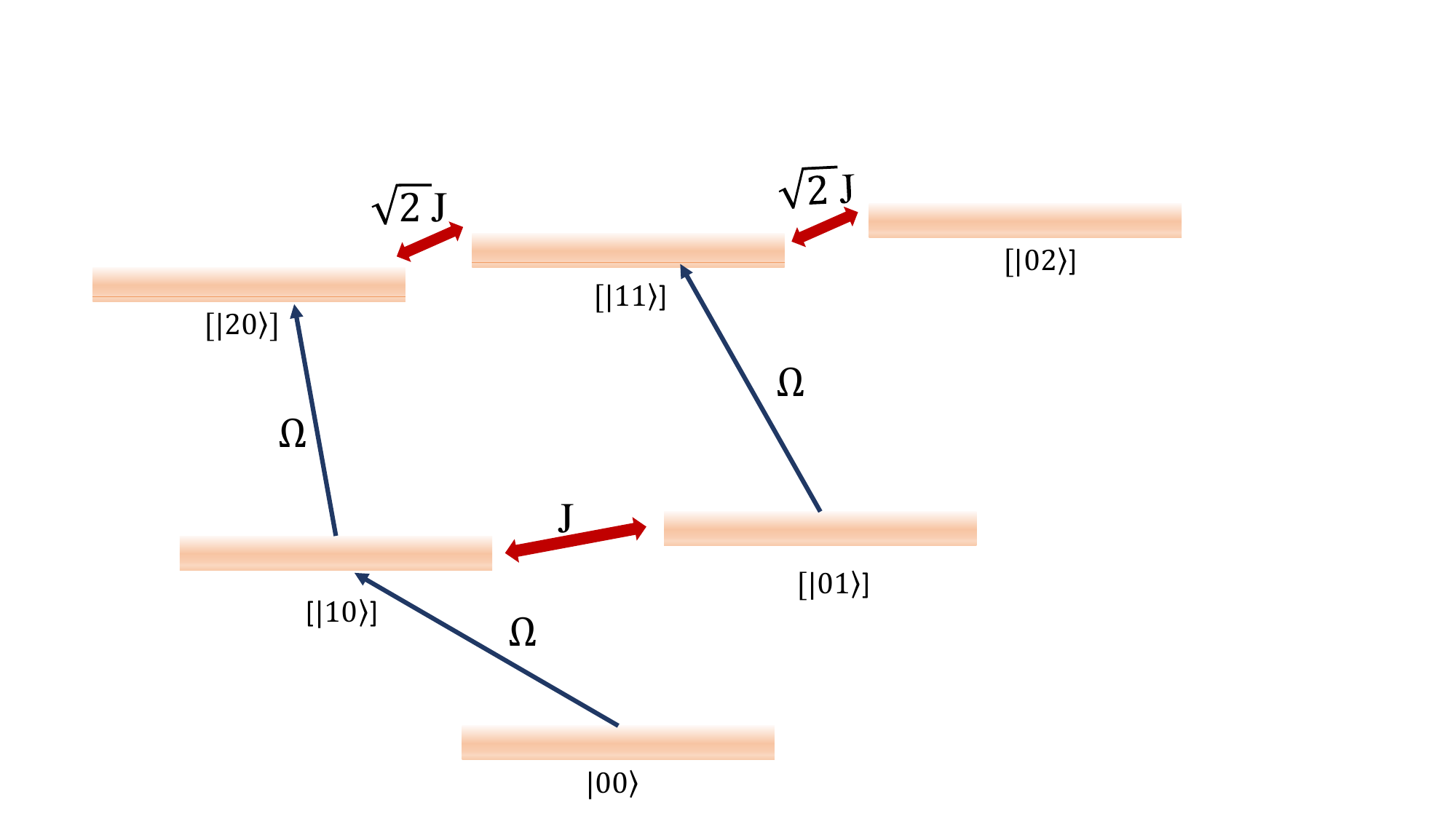}
	\caption{The eigenenergy spectrum of the coupled cavity system in the subspace with zero, one, and two excitations.}
	\label{pic27}
\end{figure}

In the domains characterized by optomechanical weak coupling (i, iii regimes), we plot second order correlation function $g_c^{(2)}(0)$ versus $\Delta_c$ for $J=25$ meV, $g=20$ meV and $J=40$ meV, $g=13$ meV in Fig. \ref{pic12}(1a) and Fig. \ref{pic12}(2a), respectively. Here, the analytical results (solid line) is compared with the numerical results (dashed line), where there are almost matched. As it depicted, there is a dip  with value of $g_c^{(2)}(0)\approx 0.7$ and $g_{\text{photon-like}}^{(2)}(0)\approx 0.02$ due to photon blockade effect. We will explain that these dips occur because of a destructive interference between two different quantum pathway transitions which called UCPB. To analyze this effect, we obtain the probabilities of one and two-photon transition through the following standard method \cite{zou2020enhancement}.\\
In the absence of cavity field driving and interaction terms, the number operator \(N = a_c^\dagger a_c + a_p^\dagger a_p\) commutes with the free Hamiltonian $ H_ 0$. We assume that the number of high-energy particles (\(p = n + m\)) in the system is either less than or equal to \(p\), while the number of phonons can be any quantity. In this context, the Hilbert space \(\mathcal{H}\) of the problem is represented as a direct sum of linear spaces \(V_i\) (\(i = 0, ..., p\)): \(V = \bigoplus_{i=0}^p V_i\), where \(V_i\) corresponds to the vector spaces of high-energy particles. The Hamiltonian restricted to the space \(V_i\) is denoted by \(H^{(i)}\), where \(i = 1, ..., p\). The eigenvectors of the Hamiltonian restricted to the space \(V\) are denoted by \(|E_{p,q}\rangle\) and can be expressed in the \(\{|nmq\rangle\}\) basis: \(|E_{p,q}\rangle = \sum_{\substack{n+m=2 \\ q}} C_{nmq} |nmq\rangle\). The indices \(p\) (\(p=n+m\)) and \(q\) correspond to the number of high-energy particles in PC-BNA cavities and the number of phonons, respectively.
Here, because of weak diving condition ($\Omega$=0.01$\kappa_{c}$) we can consider $\{C_{200}, C_{110}, C_{020}, C_{201}, C_{111}, C_{021}\} \ll \{ C_{100}, C_{010}, C_{101}, C_{011}\} \ll C_{000}$, therefore it is estimated that  \(p=2\) is a good candidate, and the Hilbert space can be decomposed as \(V=V_0 \oplus V_1 \oplus V_2\). In the \(V_0\) space, \(H^{(0)}|E_{00}\rangle = E^{(0)}_{00}|E_{00}\rangle\) with the eigenstate \(|E_{00}\rangle = |0, 0\rangle\) and the eigenvalue \(E^{(0)}_{00} = 0\). In the single-excitation subspace \(V_1\) and single phonon space, the Hamiltonian \(H^{(1)}\) is becomes:

\begin{align}
	H^{(1)}=\begin{pmatrix}
		\Delta_c & J & 0 & 0 \\
		J & \Delta_p & 0 & -g \\
		0 & 0 & \Delta_c+\omega_m & J \\
		0 & -g & J & \Delta_p+\omega_m 
	\end{pmatrix}.\label{eq12}
\end{align}
Four eigenvalues and eigenvectors can be obtained by solving the eigenvalue equation of the $H^{(1)}$ matrix. Each eigenvector $|E_{1,1}\rangle_l$ (where $l$ is the index of the eigenvector, $l=1,2,3,4$) can be expressed in the basis $\{|nmq\rangle; n+m=1\}$ as follows:

\begin{align}
	|E_{1,1}\rangle_l = f_{100}^{(1,l)}|100\rangle + f_{010}^{(1,l)}|010\rangle + f_{101}^{(1,l)}|101\rangle + f_{011}^{(1,l)}|011\rangle,\label{eq13}
\end{align}
where $f^{(1,l)}_{nmq}$ represents the components of $|E_{1,1}\rangle_l$ in the $\{|nmq\rangle; n+m=1\}$ basis.
 Similarly Hamiltonian $H^{(2)}$ in two-excitation subspace \(V_2\) and q=1 is:
\begin{align}
	H^{(2)}=\begin{pmatrix}
		\sqrt{2}J & 2\Delta_c & 0 & 0 & 0 & 0 \\
		\Delta_c+\Delta_p & \sqrt{2}J & \sqrt{2}J & -g & 0 & 0 \\
		\sqrt{2}J & 0 & 2\Delta_p+2\omega_m& 0 & 0 & -2g \\
		0 & 0 & 0 & \sqrt{2}J & 2\Delta_c+\omega_m & 0 \\
		-g & 0 & 0 & \Delta_c+\Delta_p+\omega_m & \sqrt{2}J & \sqrt{2}J \\
		0 & 0 & -2g & \sqrt{2}J & 0 & 2\Delta_p+\omega_m  
	\end{pmatrix}.\label{eq14}
\end{align}
Six eigenvectors and eigenvalues can be obtained by the eigenvalue equation of $H^{(2)}$ matrix. The eigenvectors $|E_{2,1}\rangle_l; l=1,2,...,6$ versus the basis $\{ |nmq\rangle; n+m=2\}$ is:
\begin{align} |E_{2,1}\rangle_l=f_{200}^{(2,l)}|200\rangle+f_{110}^{(2,l)}|110\rangle+f_{020}^{(2,l)}|020\rangle+f_{201}^{(2,l)}|201\rangle+f_{111}^{(2,l)}|111\rangle+f_{021}^{(2,l)}|021\rangle.\label{eq15}
\end{align}
 The state vector $|\psi(t)\rangle$ in the space $V=V_0\oplus V_1 \oplus V_2 $ is:
\begin{align}
|\psi(t)\rangle=D_{00}|E_{0,0}\rangle+\sum_{i=1}^{4}D_{1i}|E_{1,1}\rangle_i+\sum_{j=1}^{6}D_{2j}|E_{2,1}\rangle_j.\label{eq16}
\end{align}
Eq. (\ref{eq16}) is rewritten in the $\lbrace|nmq\rangle \rbrace$ basis, and the probability of the state $|nmq\rangle$ is obtained as a function of the coefficients $D_{ij}$ and $f_{nmq}$. The occupation probability of state  $|000\rangle$ ($P_{|000\rangle}$) is nearly equal to one, indicating that almost the system is in this state. By manipulate the system, the probability of transition to either the one-photon or two-photon state changed. The states with given photon number ($n$) and plasmon number ($m$), and different phonon numbers ($q_i$), are in the same equivalence class ($|nmq_1\rangle \sim |nmq_2\rangle$) and their equivalence class is denoted by $[ |nm\rangle ] = \{ |nmq\rangle; q\in \mathbb{N} \}$. The occupation probability of an equivalence class is the sum of the occupation probabilities of all their elements. Here, we consider only zero and one phonon ($q=0,1$). Each class represents a group of states around the $p=n+m, q=0$ state. The one-photon occupation probability can be computed as follows:
\begin{align}
\nonumber&P_{|100\rangle}=
\sum_{i=1}^{4}\left| D_{1i}f_{100}^{1,i}\right|^2+\sum_{\substack{i,j=1\\i\neq j}}^{4} D_{1i}D_{1j}^{*}f_{100}^{1,i}f_{100}^{1,j *}\\
&P_{|101\rangle}=\sum_{i=1}^{4}\left| D_{1i}f_{101}^{1,i}\right|^2+\sum_{\substack{i,j=1\\i\neq j}}^{4} D_{1i}D_{1j}^{*}f_{101}^{1,i}f_{101}^{1,j *} \label{eq17}\\
\nonumber&P_{[|10\rangle]}=P_{|100\rangle}+P_{|101\rangle} 
\end{align}
As well as, two-photon occupation probability can be express as:
\begin{align}
	\nonumber&P_{|200\rangle}=\sum_{i=1}^{6}\left| D_{2i}f_{200}^{2,i}\right|^2+\sum_{\substack{i,j=1\\i\neq j}}^{6} D_{2i}D_{2j}^{*}f_{200}^{2,i}f_{200}^{2,j *}\\
	&P_{|201\rangle}=\sum_{i=1}^{6}\left| D_{2i}f_{201}^{2,i}\right|^2+\sum_{\substack{i,j=1\label{eq18}\\i\neq j}}^{6} D_{2i}D_{2j}^{*}f_{201}^{1,i}f_{201}^{1,j *}\\
	\nonumber&P_{[|20\rangle ]}=P_{|200\rangle}+P_{|201\rangle} 
\end{align}
 In Fig. \ref{pic12}-(1b) and \ref{pic12}-(3b), we depict the occupation states of the bare states $P_{[|10\rangle]}$ and $P_{[|20\rangle]}$ as functions of the $\Delta_c$, with the weak coupling parameter $g=20$ meV and $g=13$ meV, respectively. The solid lines (red color for $P_{[|10\rangle]}$ and orange color for $P_{[|20\rangle]}$) represent the probability when accounting for the interference terms, whereas the dashed lines (blue color for $P_{[|10\rangle]}$ and green color for $P_{[|20\rangle]}$) represent the probability when not considering the interference terms. A dip appears as a result of the quantum interference effect influencing the state transitions induced by the driving. This dip signifies the occurrence of destructive interference, indicating unconventional photon blockade.
To establish the condition for the sub-Poissonian statistics of light, where $g^{(2)}(0) \ll 1$, it is necessary that the numerator of Eq. (\ref{eq2}) be almost zero ($|C_{200}| + |C_{201}| \approx 0$). The Kerr approximation has been also proposed since the optomechanical coupling is very smaller than mechanical frequency ($g \ll \omega_{m}$) \cite{wang2020photon}. In this approximation, the coupling coefficient $g$ and Kerr parameter $K$ are related by $K = \frac{g^2}{\omega_{m}}$, and the numerical solutions of $P_{[|20\rangle]} = 0$ yield results that closely match the approximation solution.\\
Quantum interference occurs between the direct transition $[|00\rangle] \xrightarrow{\Omega} [ |10\rangle ] \xrightarrow{\Omega} [|20\rangle ]$ and the indirect transition $[|00\rangle] \xrightarrow{\Omega} [|10\rangle] \xrightarrow{J} [|01\rangle] \xrightarrow{\Omega} [|11\rangle] \xrightarrow{\sqrt{2}J} [|20\rangle]$. The class of states and corresponding transitions are presented in Fig. \ref{pic27}. The problem has a strong resemblance to the coupling of a linear cavity to a Kerr nonlinearity cavity \cite{zou2020enhancement}. In this context, the behavior of the molecular optomechanical system closely resembles the Kerr effect, with the exception of the energy levels corresponding to molecular phonons around the optical and plasmonic cavity levels, which distinguish and group them accordingly. The cross-product terms on the right-hand side of Eq. (\ref{eq17}) and Eq. (\ref{eq18}) represent the interference probability term, which plays a crucial role in determining the UCPB mechanism. For a large PC-BNA coupling $J$, the ${|n_{+} n_{-} q\rangle}$ basis can be employed to determine the path of quantum interference and the PB effect. The class of states $[ |n_{+} n_{-}\rangle ]$ and corresponding transitions are similar to those presented in Fig. \ref{pic27}, with the replacement of $n$ by $n_{+}$ and $m$ by $n_{-}$.

\subsection{strong coupling regime}
\begin{figure}[htbp]
	\centering
	\includegraphics[width=0.4\textwidth]{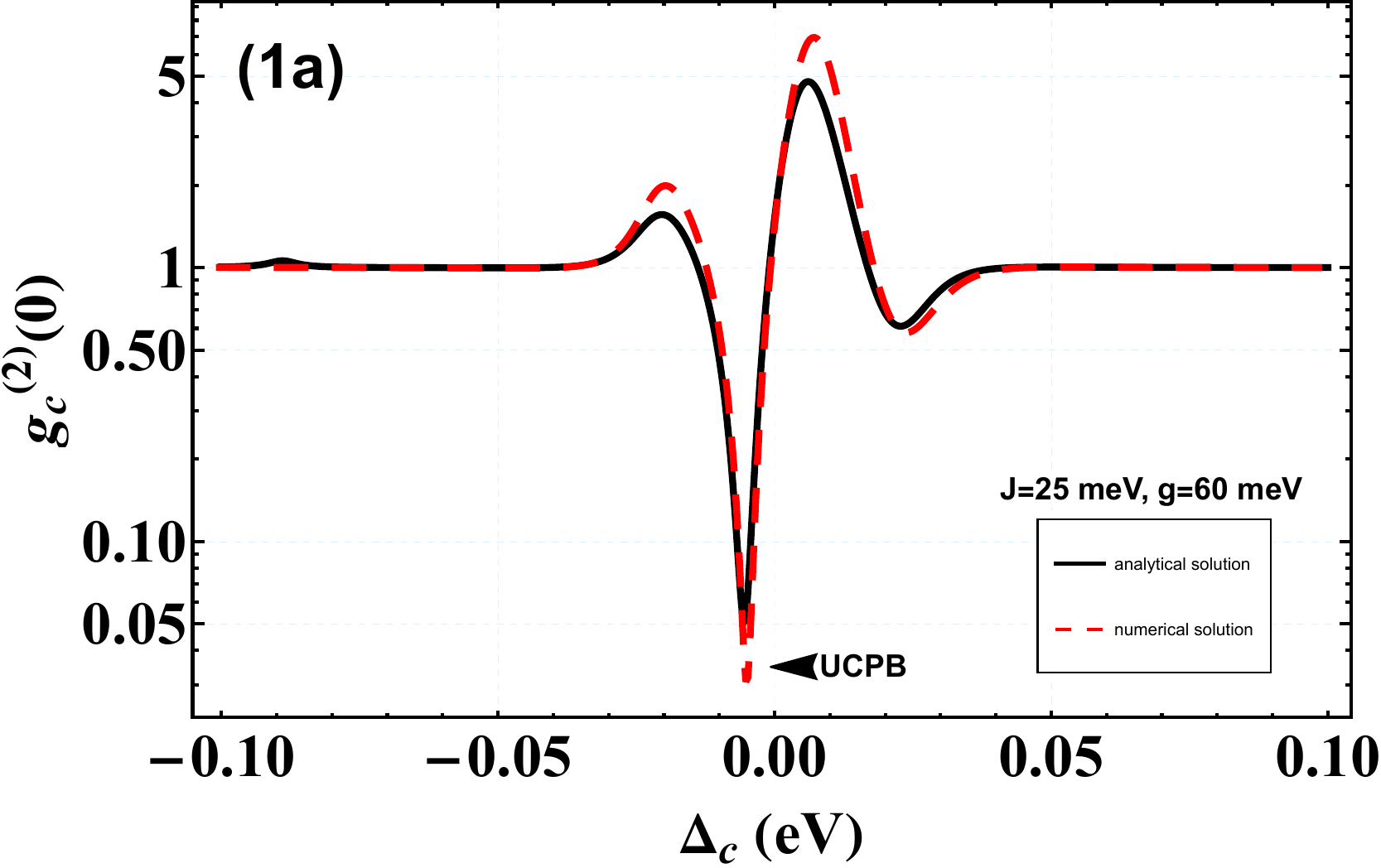}
	\hspace{1cm}
	\includegraphics[width=0.4\textwidth]{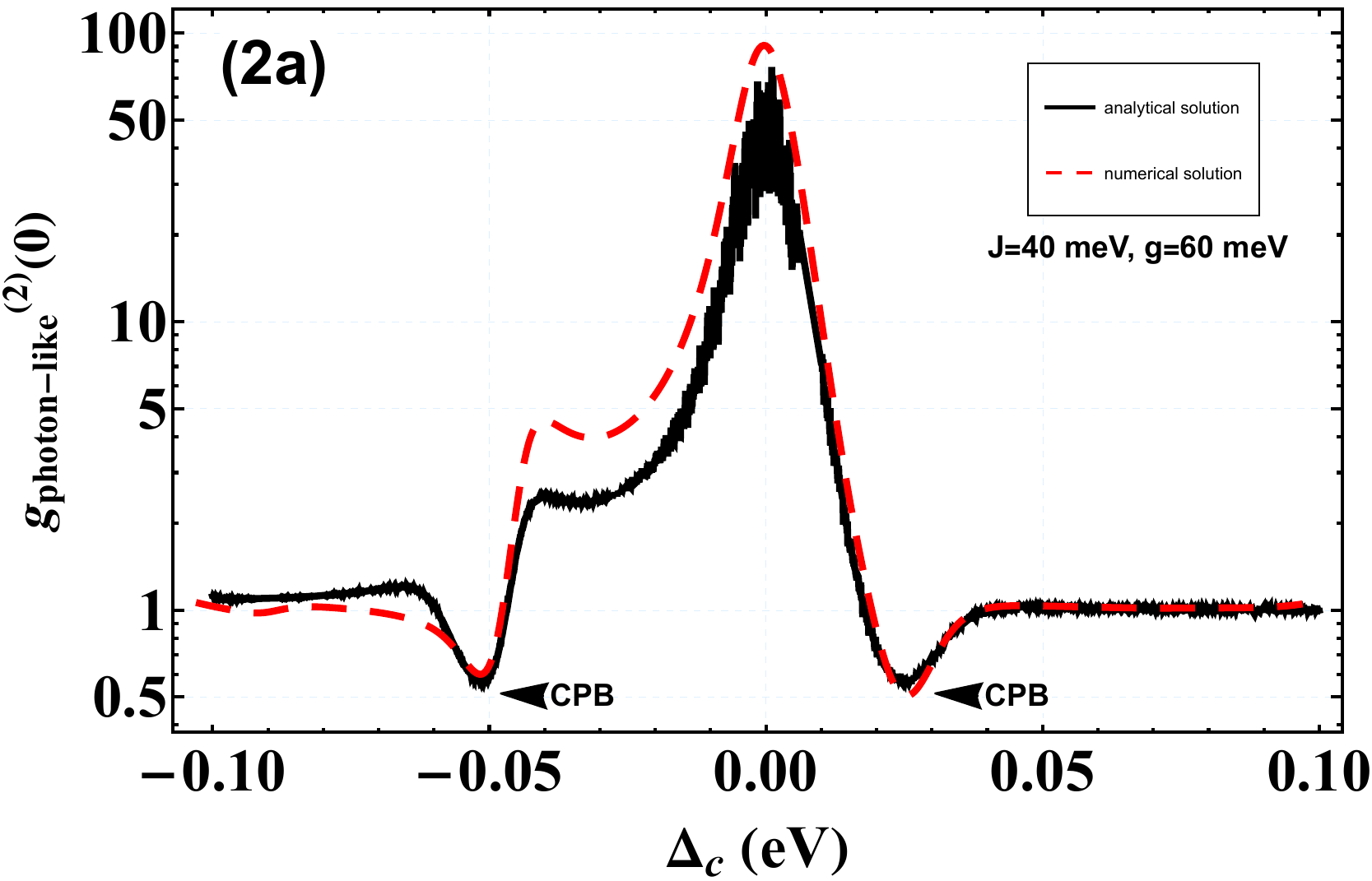}
	\hspace{1cm}
	\includegraphics[width=0.41\textwidth]{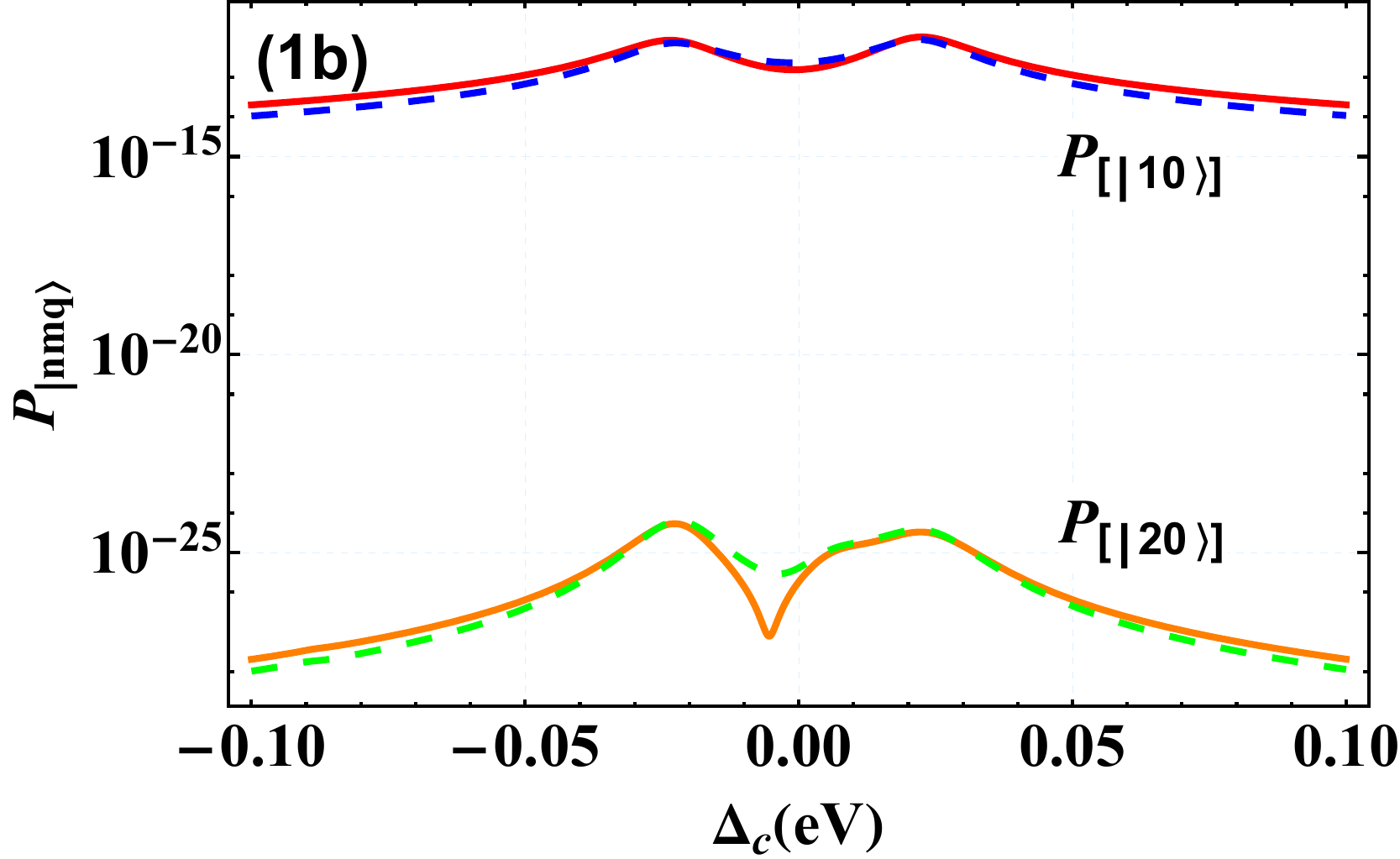}
	\hspace{0.9cm}
	\includegraphics[width=0.41\textwidth]{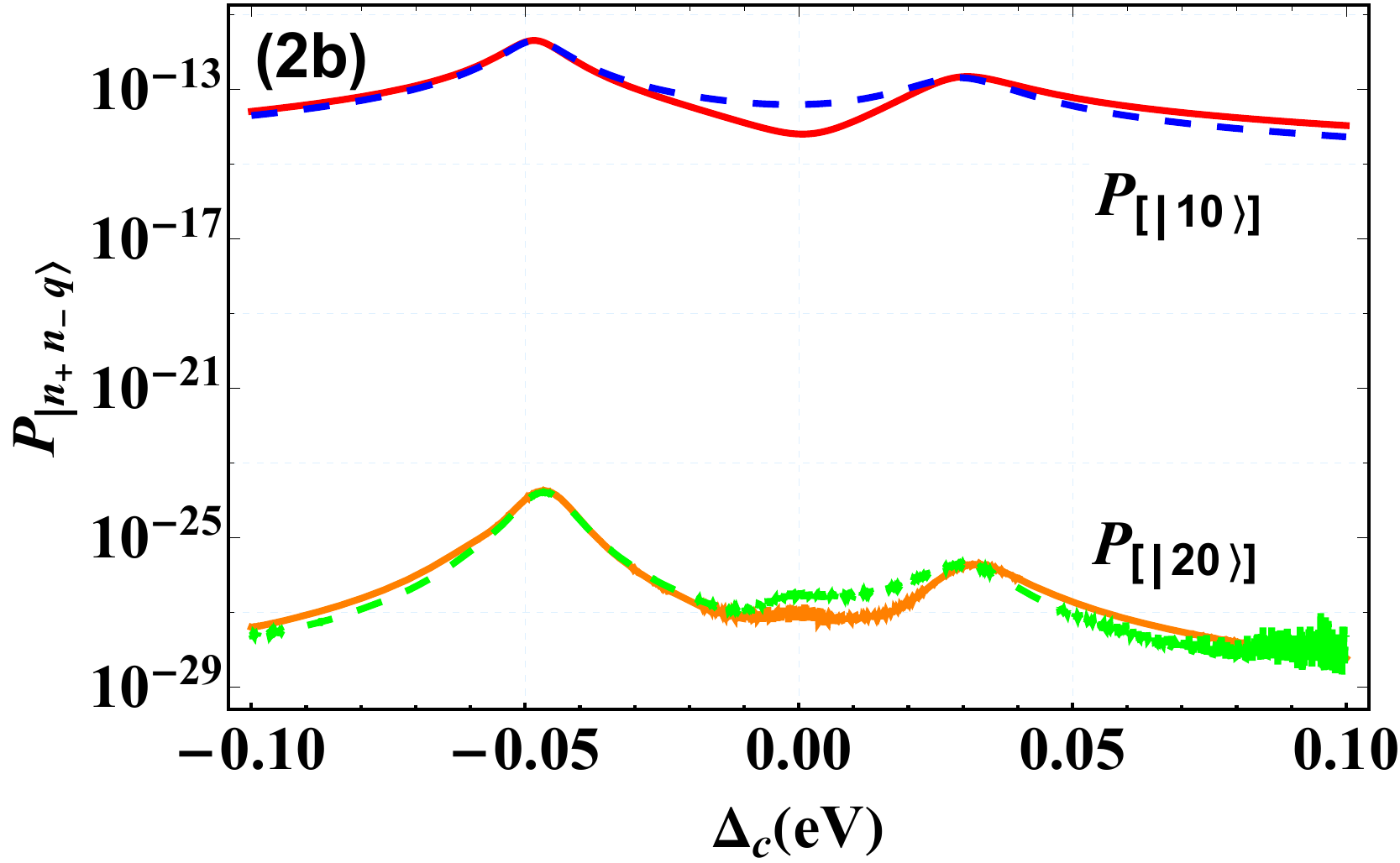}
	\caption{(1a) The correlation function $g^{(2)}(0)$ (1b) The state occupations $P_{[|10\rangle]}$ and $P_{[|20\rangle]}$ for the bare states versus the detuning $\Delta_{c}$ for $J=25$ meV and $g=60$ meV with other parameters set to $\kappa_c=1.342 \times 10^{-3}$ meV, $\kappa_p=34$ meV, $\gamma_m=200$ meV, $\omega_{c}=1.342$ eV, $\omega_{p}=1.36$ eV (3a) The correlation function $g^{(2)}(0)$. (2a) The correlation function $g^{(2)}(0)$ (2b) The state occupation probabilities $P_{[|10\rangle]}$ and $P_{[|20\rangle]}$ for the bare states versus the detuning $\Delta_{c}$ for $J=40$ meV and $g=60$ meV with other parameters set to $\kappa_c=1.32 \times 10^{-3}$ meV, $\kappa_p=27$ meV, $\gamma_m=150$ meV, $\omega_{c}=1.32$ eV, $\omega_{p}=1.36$ eV.}
	\label{pic13}
\end{figure}

In strong optomechanical coupling (ii, iv regimes), we plot second order correlation function $g_c^{(2)}(0)$ versus $\Delta_c$ for $J=25$ meV, $g=60$ meV and $J=40$ meV, $g=60$ meV in Fig. \ref{pic13}(1a) and Fig. \ref{pic13}(2a), respectively. Here, the analytical results (solid line) is compared with the numerical results (dashed line). There is a dip in the regime (ii) with value of $g^{(2)}(0)\approx 0.02$, due to UCPB photon blockade effect as it shown in Fig. \ref{pic13}-(1a), in which the probabilities of one and two photons transition are dependent to quantum interference and hence UCPB happens (see Fig \ref{pic13}-(1b)). In the strong coupling regime (iv), there are two dips because of CPB mechanism as it depicted in Fig. \ref{pic13}-(2a). Here, the probabilities of one and two photons are independent of quantum interference (dips at $\Delta_{c}=-0.05$ eV and $\Delta_{c}=0.025$) as it  shown in Fig. \ref{pic13}-(2b). 

\section{Conclusion}\label{Conclusion}
In conclusion, we have studied the potential of an integrated  molecular optomechanical system in a hybrid cavity for room temperature single photon source. For this point, We have theoretically demonstrated and calculated the second order correlation function for different coupling regime through the definition of efficient basis. We have also investigated the probabilities interference in order to adopt each observed dips to CPB or UCPB mechanisms.  Meanwhile, $g^{(2)}(0) \approx 0.02$ is achieved even in the weak molecular optomechanics coupling which is practically applicable. Our proposed model can play a pivotal role as an integral component in photonic based quantum computer and quantum networks.

\appendix
\renewcommand*{\thesection}{Appendix~\Alph{section}}
\section{}\label{appendA}
Substituting the non-Hermitian Hamiltonian given by Eq. (\ref{eq3}) and the system states from Eq. (\ref{eq4}) into the Schrödinger equation, we obtain a set of linear differential equations for the probability amplitudes, as follows:
\begin{align}
		-i\dot{C}_{100} &= (\Delta_c - i\frac{k_c}{2})C_{100} + JC_{010} + \Omega(1 + \sqrt{2}C_{200}) \nonumber\\
		-i\dot{C}_{010} &= (\Delta_p - i\frac{k_p}{2})C_{010} + JC_{100} - gC_{011} + \Omega C_{110} \nonumber\\
		-i\dot{C}_{110} &= 	(\Delta_c - i\frac{k_c}{2})C_{110} + (\Delta_p - i\frac{k_p}{2})C_{110} + \sqrt{2} J(C_{200} + C_{020}) - gC_{111} + \Omega C_{010} \nonumber\\
		-i\dot{C}_{200} &= (2\Delta_c - i\frac{k_c}{2})C_{200} + \sqrt{2}JC_{110} + \sqrt{2} \Omega C_{100}, \nonumber\\
		-i\dot{C}_{020} &= (2\Delta_p - i\frac{k_p}{2})C_{020} + \sqrt{2}JC_{110} - 2gC_{021}\nonumber\\
		-i\dot{C}_{201}&=(2\Delta_p - i\frac{k_p}{2})C_{021} + \sqrt{2}JC_{111} - 2gC_{020} + (\omega_m - i\frac{\gamma_m}{2})C_{021}\\
		-i\dot{C}_{021}&=(2\Delta_c - i\frac{k_c}{2})C_{201} + \sqrt{2}JC_{111} + \Omega C_{101} + (\omega_m - i\frac{\gamma_m}{2})C_{201}\nonumber\\
		-i\dot{C}_{111}&=(\Delta_c - i\frac{k_c}{2})C_{101} + (\omega_m - i\frac{\gamma_m}{2})C_{101} + JC_{011} + \Omega C_{111} \nonumber\\
		-i\dot{C}_{011}&=(\Delta_p - i\frac{k_p}{2})C_{011} + (\omega_m - i\frac{\gamma_m}{2})C_{011} + JC_{101} - gC_{010} + \Omega C_{111} \nonumber \\
		-i\dot{C}_{101}&=(\Delta_c - i\frac{k_c}{2})C_{111} + (\Delta_p - i\frac{k_p}{2})C_{111} + (\omega_m - i\frac{\gamma_m}{2})C_{111} \nonumber \\&+ \sqrt{2}J(C_{021} + C_{201}) - gC_{110} + \Omega C_{011}\nonumber \label{eq19}
\end{align}
To study the steady-state photon statistical properties of the system, we can disregard changes in amplitude over time.

\begin{backmatter}
	\bmsection{Funding}
	\smallskip
	\bmsection{Acknowledgments}
	\bmsection{Disclosures} The authors declare no conflicts of interest.
	\bmsection{Data availability} Data underlying the results presented in this paper are not publicly available at this time but may be obtained from the authors upon reasonable request.
\end{backmatter}

\end{document}